# Systematic Noise

Micro-movements in Equity Options Markets

**Abstract**

Equity options are known to be notoriously difficult to price accurately, and even with the development of established mathematical models there are many assumptions that must be made about the underlying processes driving market movements. As such, the theoretical prices outputted by these models are often slightly different from the realized or actual market price. The choice of model traders use can create many different valuations on the same asset, which may lead to a form of systematic micro-movement or noise. The analysis in this paper demonstrates that approximately 1.7%-4.5% of market volume for options written on the SPY ETF within the last two years could potentially be due to systematic noise.

Adam Wu

Indiana University

April 2017

Special thank you to Dr. Michael Alexeev for research process guidance and support

---

107 S Indiana Ave, Bloomington IN 47405

adamwu@indiana.edu / +1 (415) 969 0380

**Introduction**

The first known usage of an option dates back several thousand years ago to the ancient Greek philosopher Thales of Miletus, as Aristotle describes in *Politics*. Predicting that the coming season would bear a very successful olive harvest, Thales, being merely a poor philosopher, acquired the *right* to use the olive presses that turned olives into oil for a small deposit. When the season came and Thales' prediction about the weather turned out to be correct, he was able to make enormous profits for a very small investment. This contract that Thales made with the owners of the olive presses is no different from the many modern financial contracts we use today, collectively known as *derivatives*.

The focus of this paper is on a special type of financial derivative known as an *option*. An option is a contract that gives the contract-holder the right (but not the obligation) to buy or sell an asset, within a specified period of time, at a specified price. To clarify some of the terminology, this specified price is called the option's *Strike Price*, and the last day that the contract may be exercised is called the *Maturity Date*.

Further, there are two main classes of options. *European* options are ones that can only be exercised on a specific date (i.e. March 25, 2019), while *American* options may be exercised at any point up to the maturity date. In recent years, there has been the rise of more complex derivatives such as Asian options, compound options (options written on an option), and others with an embedded equity or debt security—which are even more difficult to value accurately and are outside the scope of this paper. In practice, nearly all public exchange-traded stock options in the world are of the American form[1].

From the anecdote above, it's clear that options can have very lucrative payoffs while limiting downside risk. However, an important question is on the actual valuation of these options due to their complexity (i.e. how much should Thales actually pay the olive press owners?). In the early 1970s, a major breakthrough in pricing equity options was achieved by mathematicians known as the Black-Scholes model. Since then, the Black-Scholes model has been extended on heavily, with many of its assumptions challenged and relaxed, leading to a number of other models such as analytical approximations, lattice tree methods, or statistical simulations.

However, as with any theoretical model, each comes with their own set of assumptions and computational methods which makes the pricing of an option in practice *slightly* different depending on which model one uses. As these valuations may form a key factor in trading strategies and decisions, the choice of model and their respective mispricings may be creating a form of systematic micro-movement in the equity options market.

---

[1] Note that the terms European and American do not have anything to do with the continents. The labels were coined in a 1965 article by Nobel Laureate Paul Samuelson as he was supposedly discouraged from researching options due to their complexity by his European colleagues, and so proceeded to name the simple options 'European'.

## Overview of Derivative Pricing Models

Black-Scholes-Merton Options Model

In 1973, mathematicians Fischer Black, Myron Scholes, and Robert Merton developed a stochastic differential equation that modelled the fair price of an option over time, winning the Nobel prize for economics in 1997. As one of the earliest formal mathematical models for the valuation of options, this breakthrough has had immense impact on how traders price these assets or hedge their portfolios even today. This section will *briefly* explain the main ideas behind the Black-Scholes Model, as well as cover its key assumptions and limitations[1].

Let

$$\mu = Annual\ expected\ return\ on\ underlying\ stock$$

$$\sigma = Annual\ volatility\ of\ stock$$

First, suppose that changes in stock prices are approximately normally distributed in short periods of time.

$$\frac{\Delta S}{S} \sim N(\mu \Delta t, \sigma^2 \Delta t)$$

where S is the stock price at time t and N(x,y) is a normal distribution with mean x, variance y

Assume the stochastic process for underlying price movements follow geometric Brownian motion[2].

$$dS = \mu S\ dt + \sigma S\ dz$$

Note that $\mu$ and $\sigma$ are constant, so it can be shown that (with an application of Itô's lemma):

$$\ln S_T \sim N[\ln S_0 + \left(\mu - \frac{\sigma^2}{2}\right)T, \sigma^2 T]$$

Thus, it is implied that the stock's prices are approximately log-normally distributed, at least for very short periods of time.

Now suppose that *f* is the price of a financial derivative (any asset that has its value derived from the underlying stock price). Since *f* is a function of *S* and *t*, by Itô's lemma:

$$df = \left(\frac{\partial f}{\partial S}\mu S\ dt + \frac{\partial f}{\partial t} + \frac{\partial^2 f}{\partial S^2}\frac{\sigma^2 S^2}{2}\right)dt + \left(\frac{\partial f}{\partial S}\sigma S\right)dz$$

By constructing a portfolio that shorts one derivative (-1 *f*) and long $\partial f / \partial S$ shares of the underlying stock:

$$\pi = -f + \frac{\partial f}{\partial S}S$$

where π is defined as the portfolio's value

---

[1] For a full derivation, see the authors' original work "The Pricing of Options and Corporate Liabilities." or John C. Hull's comprehensive text "Options, Futures, and Other Derivatives"

[2] This representation is widely used for modelling stock price behavior, however its actual applicability is debated and so serves as one of the main assumptions behind the Black-Scholes Model

The change of this portfolio over small periods of time is then:

$$d\pi = -df + \frac{\partial f}{\partial S} dS$$

We have already defined *df* and *dS*, and so substituting them into above gives us:

$$d\pi = \left(-\frac{\partial f}{\partial t} - \frac{\partial^2 f}{\partial S^2} \frac{\sigma^2 S^2}{2}\right) dt$$

It is important to note that the Wiener process component *dz* is cancelled out, which suggests that this portfolio is riskless during small periods of time *dt*.

$$d\pi = r\pi \, dt$$

$$\left(-\frac{\partial f}{\partial t} - \frac{\partial^2 f}{\partial S^2} \frac{\sigma^2 S^2}{2}\right) dt = r\left(-f + \frac{\partial f}{\partial S} S\right) dt$$

By rearranging terms,

$$\frac{\partial f}{\partial t} + \frac{\partial f}{\partial S} rS + \frac{\partial^2 f}{\partial S^2} \frac{\sigma^2 S^2}{2} = rf$$

Which is the full Black-Scholes-Merton differential equation that models the financial derivative's price with respect to time and the underlying stock price. Note that this has infinitely many solutions corresponding to any asset that derives its value from *S*. To solve for a specific derivative, we can define *f* with a *boundary condition*. In the case of a European call option, if *S* drops below the strike price *K*, exercising the option would result in a loss so it makes more sense to simply let it expire worthless[1].

$$f = \max(S - K, 0) \text{ at the exercise date } T$$

In addition, it is important to note that the term μ drops out of the BS differential equation. Since the term represents annual expected returns, it is affected by the individual risk preferences of investors which makes actually solving the differential equation impossible. But since the BS model is independent of risk preferences, we can impose a risk-neutral assumption (i.e. $\mu = r$).

Thus, solutions to the differential equation for European call and put options on non-dividend stocks:

$$c_t = S_t \, \phi(d_1) - Ke^{-r(T-t)} \phi(d_2)$$

$$p_t = Ke^{-r(T-t)} \phi(-d_2) - S_t \, \phi(-d_1)$$

$$\text{Where } d_1 = \frac{\ln\left(\frac{S_t}{K}\right) + \left(r + \frac{\sigma^2}{2}\right)(T-t)}{\sigma\sqrt{T-t}}$$

$$d_2 = d_1 - \sigma\sqrt{T-t}$$

$$\phi(x) = \frac{1}{\sqrt{2\pi}} \int_{-\infty}^{x} e^{-\frac{x^2}{2}} \text{ or the CDF for a standard normal distribution}$$

---

[1] As mentioned in the introduction, European call options can only be exercised at a specific time T, while American options may be exercised at any point up to then. It can be shown that while the BSM model can't directly apply to American options, its valuation can be broken down into a theoretical European price plus an early exercise premium. Thus the BSM model can be useful even for more complex derivatives by acting as a lower bound in approximations. See Black's Approximation.

Model Assumptions

To recap, the main assumptions used to derive the BSM differential equation:

- Underlying stock price follows a geometric Brownian motion with known and constant $\mu$ and $\sigma$
- Market permits short selling securities, with no transaction costs or taxes on capital gains
- Prices are continuous and so are transactions, suggesting that securities are perfectly divisible (i.e. it is possible to buy 1.0592849… amounts of shares)
- Risk-free rate is constant and the same across all derivatives with different maturities

Main assumptions for solutions of the PDE to price European options:

- Underlying stock does not pay dividends
- Risk-neutral probability measure (i.e. the stochastic process is a martingale)

Many of these assumptions of the original BSM model are challenged and can be relaxed. Even the most fundamental ones about the underlying price movements are questioned, with studies suggesting that the distribution of returns is in fact leptokurtic, skewed, and even prone to discontinuous jumps (see Anderson, et al. "The Distribution of Stock Return Volatility"). In practice, the Black-Scholes is still widely used despite its limitations given its ease of computation, though practitioners (especially at quantitative hedge funds or trading desks) would usually slightly adjust the model. This can range from simply treating the volatility and risk-free parameters as non-constant via autoregressive models, to building sophisticated stochastic-volatility jump-diffusion models as alternatives to the BS. However the BS model is still incredibly important in its role as a benchmark (and potentially valuation bounds) in the market, and it's not unreasonable that a significant number of practitioners still heavily rely on the BS as its simplicity can outweigh the slight improvement in accuracy from more complex models. This paper then focuses on two relatively simple pricing models, the Black-Scholes and the Barone-Adesi and Whaley (BAW) analytical approximation. Thus, while there are certainly other models that are used to price options, these two should capture a significant portion of any systematic market movements between them as even the more sophisticated models are only marginally more accurate in practice.

Barone-Adesi and Whaley Model

Published in 1987, just 14 years after the Black-Scholes model, researchers Giovanni Barone-Adesi and Robert Whaley developed a simple analytical approximation method to price American options (BAW). Traditional methods to price American options at the time were extremely computationally inefficient and difficult, and as is demonstrated by the BS model, there are no closed-form analytical solutions (one such method include working with binomial trees in discrete time, see Cox-Ross-Rubinstein). Another advantage is that the BAW model can be extended to price options written on dividend-paying stocks.

First, the BAW makes the same assumptions as the BS regarding the underlying price movements. Supposing that these movements can be modeled by geometric Brownian motion, Whaley claims that the partial differential equation can also apply to the early exercise premium of an American option. In other words, American options are simply European options with an added component for allowing one to exercise early.

Define the early exercise premium as

$$\varepsilon(S,T) = C(S,T) - c(S,T)$$

where C(S,T) is the American option's value and c(S,T) is the equivalent European option

Through some simplification, factoring, and re-arranging it can be shown that:

$$\varepsilon(S, X) = X(T)f(S, X)$$

$$\frac{\partial^2 f}{\partial S^2}S^2 + \frac{\partial f}{\partial S}\frac{2bS}{\sigma^2} - f\frac{2r}{\sigma^2 X} - \frac{\partial f}{\partial X}\frac{(1-X)2r}{\sigma^2} = 0$$

Which describes the value of the early exercise premium over time. For very short times to expiration, as (T-t) approaches 0, ∂f / ∂X also approaches 0. For very long times to expiration, (T-t) approaches ∞, X approaches 1. Thus, as an approximation, the last term is dropped.

$$\sim \frac{\partial^2 f}{\partial S^2}S^2 + \frac{\partial f}{\partial S}\frac{2bS}{\sigma^2} - f\frac{2r}{\sigma^2 X} = 0$$

Note that this becomes an ordinary differential equation with $f = aS^q$ and parameters a, q.

$$\frac{d^2 f}{dS^2}S^2 + \frac{df}{dS}\frac{2bS}{\sigma^2} - f\frac{2r}{\sigma^2 X} = 0$$

The general solution to the differential equation is

$$f = a_1 S^{q_1} + a_2 S^{q_2}$$

Thus, it can be shown that the price of an American call option is approximately:

$$C(S,T) = c(S,T) + (\frac{S}{S^*})^{q_2} A_2 \quad \text{when S < S*}$$

$$C(S,T) = S - K \quad \text{when S ≥ S*}$$

$$\text{where} \quad A_2 = \frac{S^*}{q_2}(1 - e^{(b-r)(T-t)}\phi[d_1(S^*)]) > 0$$

$$q_2 = \frac{1}{2}[-(\frac{2b}{\sigma^2} - 1) + \sqrt{(\frac{2b}{\sigma^2} - 1)^2 + \frac{8r}{X\sigma^2}}] > 0$$

*b* is the cost of carry (b = r for non-dividend stocks, otherwise b = r − d where *d* is dividend yield)

S*, S** are critical points that can be determined iteratively[1]

Similarly, to derive the price of an American put option:

$$P(S,T) = p(S,T) + (\frac{S}{S^{**}})^{q_1} A_1 \quad \text{when S > S**}$$

$$P(S,T) = K - S \quad \text{when S ≥ S*}$$

$$\text{where} \quad A_1 = -\frac{S^{**}}{q_2}(1 - e^{(b-r)(T-t)}\phi[-d_1(S^{**})]) > 0$$

$$q_1 = \frac{1}{2}[-(\frac{2b}{\sigma^2} - 1) - \sqrt{(\frac{2b}{\sigma^2} - 1)^2 + \frac{8r}{X\sigma^2}}] < 0$$

---

[1] The critical points S*, S** can be determined by an algorithm that the authors suggested in their original work, "Efficient Analytic Approximation of American Option Values", 1987.

## The Theory

As with any financial market, option price movements are going to be *primarily* driven by expectations of the future. Thus, the current market price can be essentially viewed as a collective weighted average of future expectations. However, there has been growing dissent among financial economists as to whether markets are efficient, the existence of excess risk-adjusted returns, whether prices actually follow a random walk, and so on. This has likely contributed to the growing fragmentation of the financial industry in recent years, for example a strong believer in market efficiency might take on a more passive, index-based market approach while an active manager might try to capitalize on short-term market inefficiencies. Of course, few would believe the markets to be entirely efficient nor inefficient, and most people in fact would lie somewhere in between.

The key premise behind this paper is to recognize that there exists at least some degree of market inefficiency, no matter how large or small, and to try to isolate the systematic market movements that are caused by the existence of different valuations on the same asset. These different valuations, as discussed in the previous section, are potentially due to the usage of different pricing models in practice. Thus, it is possible that the existence of these different valuations then creates a form of systematic movement which is independent of risk preferences or expected returns.

### A Simple Demonstration

Suppose we have a theoretical market with an asset currently trading for $5.00 at $t_1$. At some time in the future $t_2$, assume that a fair valuation is $7.00. In this market, information "trickles down" in the sense that prices change gradually, where it initially moves from investors trading private information or unusual analytical ability, followed by the public who gradually adjust market prices as access to that information or analysis becomes more publicly available[1].

Thus in this very simple theoretical market, the price trend may look something like below. After the change, the price stabilizes at $7 under no-arbitrage conditions.

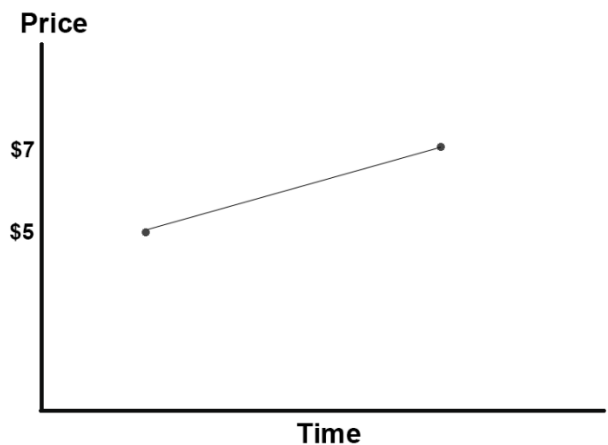

(Figure 1)

---

[1] Otherwise, without this trickle-down effect, markets would be theoretically discontinuous and immediately adjust. Whether this simple model is an accurate representation of real financial markets isn't the point, the purpose is to illustrate the effect that the existence of different valuations may have.

Now suppose that the future price at $t_2$ is unknown, and so must be estimated via a model. Let's say the market consists of two different groups of investors. Based on their fundamental views, experience, or preference the first group decides to use Model A to price this asset. The second group prefers Model B and thus prices the asset slightly differently due to the different assumptions in the two models.

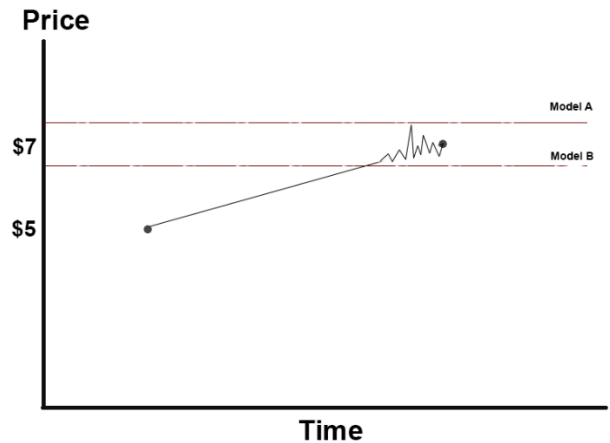

(Figure 2)

Because there are now two different valuations of the same asset, the market is unable to settle on any one specific price as it did in the previous example. At any given price, the two groups hold different fundamental views on the asset's fair valuation. If the asset's price is close to the bottom red line, Group B may find that it is a fair price however Group A thinks it is undervalued according to their model. This would lead to an increase in net long positions, driving the price up. But when the price moves up, Group B then thinks it is overvalued and begins shorting the market, driving the price back down and repeating *ad infinitum* – creating systematic up and down movements in the form of noise. In the context of the two models discussed earlier, it can be seen that the Black-Scholes price is always lower than the Barone-Adesi and Whaley approximation and serves as a lower bound.

This logic can be extended more generally to any movement in any financial market, as individual investors can have different expectations based on their access to information, risk preferences, analytical ability, or other qualitative attributes. However, the equity options market is a particularly unique application of this logic since there exists standardized ways to price options via widely accepted models, in which case a major determinant for these systematic up-and-down movements in the short-run could simply be the result of the usage of different models.

## The Empirical Model

According to the theoretical ideas behind this paper, the existence of different valuations should have a positive impact on trading volume, and this will be the main hypothesis that will be tested throughout the analysis.

To begin, recall an earlier claim that price movements are primarily driven by expectations of the future. It then follows that trading volume can be thought of as driven by the difference between current price and future expectations, such that if expectations of future price is equivalent to current price, there would be no need for any market adjustment.

And so a possible model for volume, weighted by each investor's individual capital allocation could be:

$$Volume_t = [Weighted\ Differences\ Between\ Future\ Expectations\ and\ Current\ Price] + \varepsilon_t$$

where $\varepsilon_t$ is the error term

The $\varepsilon_t$ can be interpreted as a measure of market inefficiency, which will be referred to as *noise*. We can then break this term down into systematic and non-systematic components:

$$\varepsilon_t = [Systematic\ Noise] + [Nonsystematic\ Noise]$$

Where systematic noise, as described earlier, could at least be partially due to the usage of different pricing models in practice. Non-systematic noise on the other hand is quite clearly impossible to model. For example, if someone decides to log onto his brokerage account and start randomly hitting keys on his keyboard. Thus, non-systematic noise can be interpreted both as a measure of general irrationality[1] or trades that are made without regard to expectations of future price (i.e. as part of a hedging strategy).

Denote the market option price at time t by $P_t$ and the models' theoretical price by $E[P_t \mid BS]$ and $E[P_t \mid BAW]$ such that the pricing errors are:

$$\begin{cases} X_{BS_t} = E[P_t \mid BS] - P_t \\ X_{BAW_t} = E[P_t \mid BAW] - P_t \end{cases}$$

Which gives us

$$[Systematic\ Noise] = X_{BS_t} + X_{BAW_t} + b_t$$

where $b_t$ contains other potential sources of systematic noise (ie other pricing models)

Such that

$$Volume_t = [Weighted\ Diff.\ in\ Expectations] + X_{BS_t} + X_{BAW_t} + z_t$$

---

[1] Due to the relatively complex nature of options, it isn't unreasonable to think that there is far less room for irrationality than other traded securities as the complexity essentially serves as a barrier of entry into the market.

Clearly the first term, weighted differences between expectations of the future and current price, is unobserved and impossible to model as that'll require knowledge of every investor's individual expectations, risk preferences, and capital allocation. However, from the underlying theory behind this paper, if we assume that price changes are gradual and continuous, it is possible that past volume could serve as a proxy for this first term through the momentum effect of stock behavior[1].

The autoregressive portion of the model with k-lags (k to be determined empirically later):

$$Volume_t = \varphi_0 + \varphi_1 Volume_{t-1} + \cdots + \varphi_k Volume_{t-k} + a_t$$

where $a_t$ is the white noise component of the time series

Here the term $a_t$ can be interpreted as an *innovation* at time t, referring to some new information at time t that affects the time series. Because the hypothesis is that pricing errors may be contributing to systematic market movements, these errors may then be considered partial innovations at time t.

Therefore, the theoretical model that will be tested is:

$$Volume_t = \varphi_0 + \varphi_1 Volume_{t-1} + \cdots + \varphi_k Volume_{t-k} + \varphi_{k+1} X_{BS_t} + \varphi_{k+2} X_{BAW_t} + u_t$$

where $u_t$ is a white noise series[2]

---

[1] Momentum is the idea that price movements may follow a trend over time rather than instantly reflecting new information. This phenomenon is increasingly supported by literature and even accepted by Eugene Fama of the Efficient Market Hypothesis as a market anomaly.

[2] Note that the term *white noise* is the time series' equivalent of an error term, and not to be confused with the paper's definition of financial noise as systematic market movements

**The Methodology**

Description of the Data

End-of-day (EOD) historical options data written on the SPDR S&P 500 Trust ETF (SPY) over the last two years (01/02/2015 – 12/30/2016) was collected from data service IVolatility. As the SPY is one of the most actively traded ETFs in the world and also holds one of the deepest options markets, this makes it a good candidate for analysis as the underlying theoretical ideas implicitly assume an actively traded and liquid market. In addition, the widely-diversified nature of an index-tracking ETF makes it less prone to sudden price shocks, such as unexpected news or announcements or legislature that would affect an individual company or industry more severely. The relatively smoother movements of the SPY are also more in line with the continuous price trend assumptions described earlier.

**Descriptive Statistics of Options written on SPY (EOD) from 01/02/15 - 12/30/16**

| Statistic | N | Mean | St. Dev. | Min | Max |
|---|---|---|---|---|---|
| adjusted.stock.close.price | 2,089,590 | 207.806 | 8.644 | 182.860 | 227.760 |
| strike | 2,089,590 | 190.159 | 53.474 | 7.000 | 350.000 |
| ask | 2,089,590 | 22.561 | 34.050 | 0.000 | 219.500 |
| bid | 2,089,590 | 22.164 | 33.631 | 0.000 | 217.090 |
| volume | 2,089,590 | 622.227 | 4,196.244 | 0 | 667,769 |
| open.interest | 2,089,590 | 4,902.684 | 16,783.390 | 0 | 565,321 |

The data is then filtered for existing trading activity as any analysis on contracts that aren't traded or don't actually exist is fairly meaningless and will skew the results[1]. Thus, any interpretation moving forwards should take into account that the data was filtered for non-zero volume and open interest which limits the dataset to only actively traded options contracts between 2015-2016.

**Descriptive Statistics of Actively Traded SPY Options**

| Statistic | N | Mean | St. Dev. | Min | Max |
|---|---|---|---|---|---|
| adjusted.stock.close.price | 705,885 | 208.023 | 8.912 | 182.860 | 227.760 |
| strike | 705,885 | 196.642 | 29.455 | 8.000 | 350.000 |
| ask | 705,885 | 6.787 | 11.822 | 0.000 | 201.290 |
| bid | 705,885 | 6.622 | 11.620 | 0.000 | 199.710 |
| volume | 705,885 | 1,495.250 | 5,940.193 | 1 | 500,656 |
| open.interest | 705,885 | 11,118.550 | 24,974.550 | 1 | 565,321 |

---

[1] Note that at any given time, there can be a large number of possible variations of options contracts with different maturities or strikes, but only a few dozen are actually actively traded. *Open Interest* is a measure of the number of open contracts currently on the market, and so the data is filtered for activity by imposing the condition that open interest and volume is non-zero.

In addition, treasury rates and historical volatility of the underlying index were used as parameters in the computation of theoretical prices.

**Average Treasury rates from 2015-2016**

| Statistic | N | Mean | St. Dev. | Min | Max |
|---|---|---|---|---|---|
| X1.mo | 501 | 0.145 | 0.125 | 0.000 | 0.500 |
| X3.mo | 501 | 0.186 | 0.154 | 0.000 | 0.550 |
| X6.mo | 501 | 0.314 | 0.184 | 0.050 | 0.660 |
| X1.yr | 501 | 0.468 | 0.189 | 0.160 | 0.920 |
| X2.yr | 501 | 0.760 | 0.160 | 0.440 | 1.290 |
| X3.yr | 501 | 1.014 | 0.171 | 0.660 | 1.610 |
| X5.yr | 501 | 1.433 | 0.228 | 0.940 | 2.100 |
| X7.yr | 501 | 1.763 | 0.259 | 1.190 | 2.420 |
| X10.yr | 501 | 1.988 | 0.277 | 1.370 | 2.600 |
| X20.yr | 501 | 2.383 | 0.298 | 1.690 | 2.980 |
| X30.yr | 501 | 2.718 | 0.275 | 2.110 | 3.250 |

**Historical Volatility of Underlying SPY ETF**

| Statistic | Mean | St. Dev. | Min | Max |
|---|---|---|---|---|
| Volatility | 0.132 | 0.061 | 0.044 | 0.380 |

Computation of Theoretical Prices

The theoretical prices outputted by the Black-Scholes and Barone-Adesi and Whaley models were computed for the filtered dataset using an R script, using a collection of functions within the package fOptions. As defined previously, the pricing errors are the difference between the model's expected price and the market's current price (mid-point between bid and ask).

**Pricing Errors[1] ($)**

| Statistic | N | Mean | St. Dev. | Min | Max |
|---|---|---|---|---|---|
| BS.Error | 1,458,867 | -0.978 | 2.708 | -22.956 | 24.973 |
| BAW.Error | 1,458,867 | -0.786 | 2.800 | -22.402 | 24.989 |

As seen from the output, both models tend to under-estimate the option price which is an observed phenomenon in literature as implied volatilities tend to be higher than historical volatilities. Some researchers point to behavioral economics in which investors tend to demand a risk premium as compensation rather than the risk-neutrality that the models assume (see Volatility Smile).

[1] A trailing 21-day historical volatility was used for the computations of the model prices across options with different strikes and maturities. While this is consistent with the Black-Scholes assumption of a constant volatility term structure, in practice it may be more precise to use historical volatility rates that scale with time to expiration. As a result of this assumption, a handful of theoretical prices were excessively large (near infinite) and so a handful of outliers above 25 (9 s.d. away) were dropped.

## The Analysis

From before, the theoretical model to be tested is:

$$Volume_t = \varphi_0 + \varphi_1 Volume_{t-1} + \cdots + \varphi_k Volume_{t-k} + \varphi_{k+1} X_{BS_t} + \varphi_{k+2} X_{BAW_t} + u_t$$

where $u_t$ is a white noise series

As this is a k-lagged autoregressive model, a important procedure is to ensure that the time series is *stationary*. Any statistical inference in time series analysis heavily relies on an assumption of weak stationarity, which means that the time series' first two moments, mean and variance, do not vary over time. In addition, a further assumption that will be important for prediction is that these two moments are also finite.

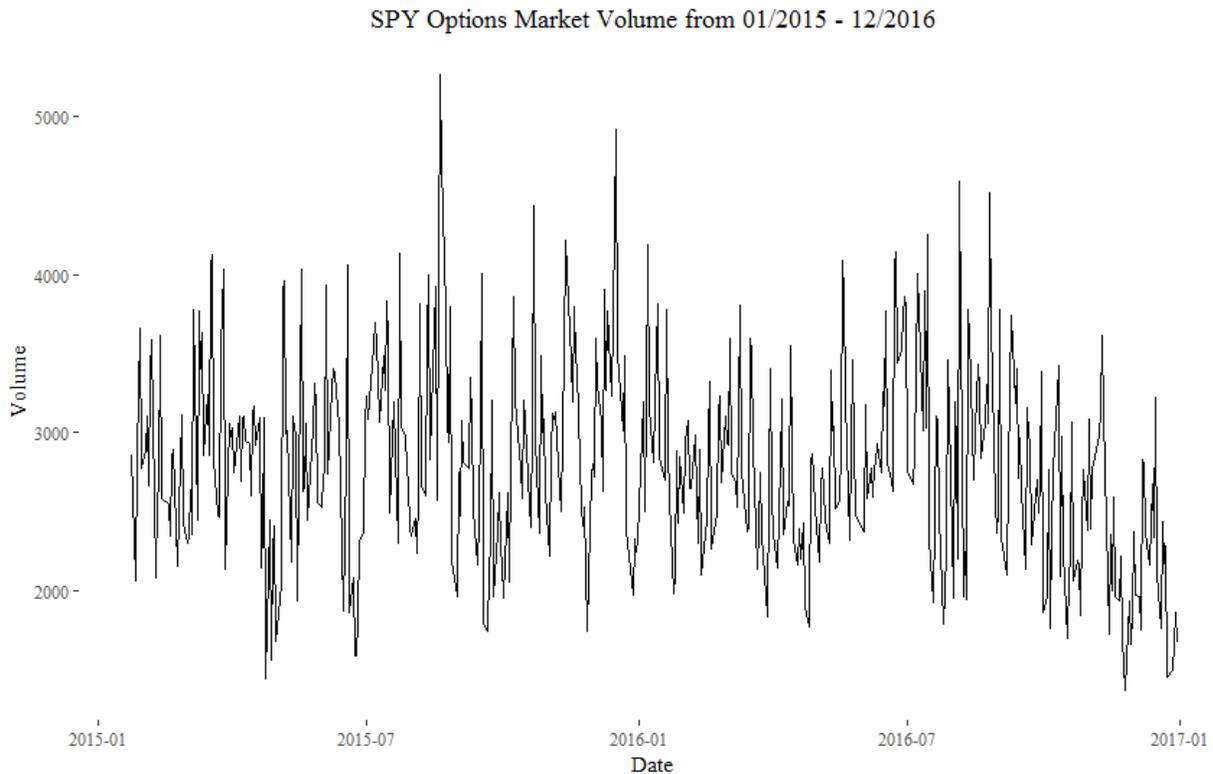

(Figure 3)

| Statistic | N | Mean | St. Dev. | Min | Max |
| --- | --- | --- | --- | --- | --- |
| Volume | 486 | 2,820.531 | 635.295 | 1,367.693 | 5,268.078 |

(Figure 4)

From Figure 3, the market volume for SPY options seems to stay around the mean, and the variance is roughly constant over time, so the time series appears to be approximately stationary. To formally test this, one method is the *augmented Dickey-Fuller* test to verify the existence of a unit-root. But first, we have to determine the k number of lags to use.

Identifying the Order of the AR Model

In order to determine an appropriate time lag to use for the autoregressive portion of the model (past volume), one approach is to use the partial autocorrelation function (PACF)[1].

First, consider the following AR models of varying orders in succession:

$$Volume_t = \varphi_{0,1} + \varphi_{1,1} Volume_{t-1} + \epsilon_{1,t}$$

$$Volume_t = \varphi_{0,2} + \varphi_{1,2} Volume_{t-1} + \varphi_{2,2} Volume_{t-2} + \epsilon_{2,t}$$

$$Volume_t = \varphi_{0,3} + \varphi_{1,3} Volume_{t-1} + \varphi_{2,3} Volume_{t-2} + \varphi_{3,3} Volume_{t-3} + \epsilon_{3,t}$$

$$\ldots$$

$$Volume_t = \varphi_{0,j} + \varphi_{1,j} Volume_{t-1} + \cdots + \varphi_{j,j} Volume_{t-j} + \epsilon_{j,t}$$

Let $\hat{\varphi}_{1,1}$ denote the sample coefficient on the first time-lag, formally called the *lag-1 sample PACF*. Notice that these models are multiple linear regressions and can be estimated by OLS. Thus, similar to the logic behind partial F tests, we can find the value of $\hat{\varphi}_{j,j}$ after which consecutive coefficients are insignificant.

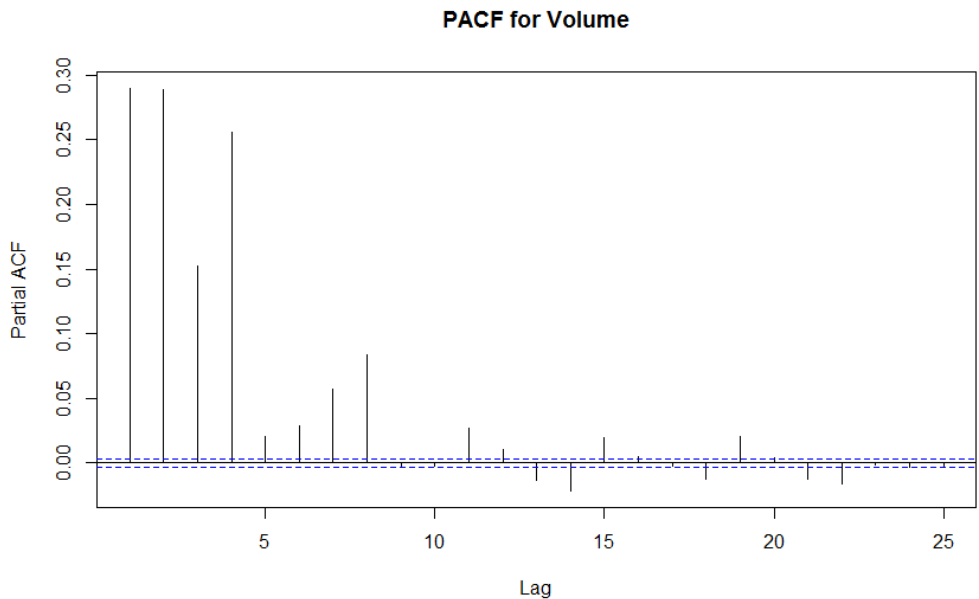

(Figure 5)

The blue lines mark significance at the 95% level, which suggests that an autoregressive model with order 22 (i.e. 22 time lags) may be appropriate. Here it should be noted that the sample PACF does decay exponentially rather rapidly, which again supports the stationarity claim earlier.

---

[1] Other methods include informational criteria methods such as the Akaike and Schwarz-Bayesian criterion, but they give similar results and the PACF may be more intuitive

Testing for Stationarity

The augmented Dickey-Fuller (ADF) tests whether a time series has a unit root or equivalently, follows a random walk[1]. This is done by fitting a model of the form:

$$\Delta V_t = a_0 + \beta V_{t-1} + \delta t + a_1 \Delta V_{t-1} + \cdots + a_{22} \Delta V_{t-22} + u_t$$

The terms $a_0$ and $\delta t$ are the model's constant and drift terms for the random walk. However from Figure 3, it appears that the time series seems to vary around the mean with no visible time trend. Thus, we can drop the drift term and test the following:

$$\Delta V_t = a_0 + \beta V_{t-1} + a_1 \Delta V_{t-1} + \cdots + a_{22} \Delta V_{t-22} + u_t$$

The null and alternative hypotheses of the ADF t-test are:

$$H_0 : \beta = 0$$

$$H_A : \beta < 0$$

Or in other words, the ADF tests the null hypothesis is that Volume follows a random walk without drift with the alternative being stationarity.

```
                    Title:
            Augmented Dickey-Fuller Test

                 Test Results:
                   PARAMETER:
                  Lag Order: 22
                   STATISTIC:
             Dickey-Fuller: -121.8869
                    P VALUE:
                     0.01
```

Thus, the null hypothesis is rejected in support of the alternative hypothesis that Volume is stationary.

And with the stationarity condition fulfilled, this suggests a linear autoregressive model with order 22 and pricing errors as shocks.

$$V_t = \varphi_0 + \sum_{i=1}^{22} \varphi_i V_{t-i} + \varphi_{23} X_{BS_t} + \varphi_{24} X_{BAW_t} + u_t$$

where t = 23, …, T

---

[1] A random walk is a stochastic process that describes the path of a random variable over time. This can range from a simple set of random variables in discrete time to special cases like Brownian motion which form the basis of the Black Scholes. In the finance context, a random walk implies that stock price changes are independent and move randomly.

Model Specifications

It may be more meaningful to observe changes in volume in terms of percentages, which could justify the following logarithmic transformation:

$$\log(V_t) = \varphi_0 + \sum_{i=1}^{22} \varphi_i \log(V_{t-i}) + \varphi_{23} X_{BS_t} + \varphi_{24} X_{BAW_t} + u_t$$

In addition, it should be noted that the relationship between volume and pricing errors appears non-linear, which may further support the transformation by reducing the effect of large deviances and improving the results of the model.

The full regression output is on the following page.

**The Results**

| | Transformed Autoregressive-22 Model (Lags are logged) | | |
|---|---|---|---|
| | Effect on Market Volume (%) | | |
| | (1) | (2) | (3) |
| Black-Scholes Errors | 0.027*** | | 0.051*** |
| | (0.001) | | (0.004) |
| Barone-Whaley Errors | | 0.024*** | -0.024*** |
| | | (0.001) | (0.004) |
| Lag 1 | 0.098*** | 0.098*** | 0.098*** |
| | (0.001) | (0.001) | (0.001) |
| Lag 2 | 0.272*** | 0.272*** | 0.272*** |
| | (0.001) | (0.001) | (0.001) |
| Lag 3 | 0.082*** | 0.082*** | 0.082*** |
| | (0.001) | (0.001) | (0.001) |
| Lag 4 | 0.142*** | 0.143*** | 0.142*** |
| | (0.001) | (0.001) | (0.001) |
| Lag 5 | 0.024*** | 0.024*** | 0.024*** |
| | (0.001) | (0.001) | (0.001) |
| Lag 6 | 0.074*** | 0.074*** | 0.074*** |
| | (0.001) | (0.001) | (0.001) |
| Lag 7 | 0.013*** | 0.013*** | 0.013*** |
| | (0.001) | (0.001) | (0.001) |
| Lag 8 | 0.053*** | 0.053*** | 0.053*** |
| | (0.001) | (0.001) | (0.001) |
| Lag 9 | 0.012*** | 0.012*** | 0.012*** |
| | (0.001) | (0.001) | (0.001) |
| Lag 10 | 0.019*** | 0.018*** | 0.019*** |
| | (0.001) | (0.001) | (0.001) |
| Lag 11 | 0.006*** | 0.006*** | 0.006*** |
| | (0.001) | (0.001) | (0.001) |
| Lag 12 | 0.023*** | 0.023*** | 0.023*** |
| | (0.001) | (0.001) | (0.001) |
| Lag 13 | -0.001 | -0.001 | -0.001 |
| | (0.001) | (0.001) | (0.001) |
| Lag 14 | 0.001 | 0.001 | 0.001 |
| | (0.001) | (0.001) | (0.001) |
| Lag 15 | 0.003** | 0.003** | 0.003** |
| | (0.001) | (0.001) | (0.001) |
| Lag 16 | 0.014*** | 0.014*** | 0.014*** |

|  | (0.001) | (0.001) | (0.001) |
| --- | --- | --- | --- |
| Lag 17 | 0.004*** | 0.004*** | 0.004*** |
|  | (0.001) | (0.001) | (0.001) |
| Lag 18 | -0.005*** | -0.005*** | -0.005*** |
|  | (0.001) | (0.001) | (0.001) |
| Lag 19 | 0.005*** | 0.005*** | 0.005*** |
|  | (0.001) | (0.001) | (0.001) |
| Lag 20 | 0.006*** | 0.006*** | 0.006*** |
|  | (0.001) | (0.001) | (0.001) |
| Lag 21 | 0.002 | 0.002 | 0.002 |
|  | (0.001) | (0.001) | (0.001) |
| Lag 22 | -0.014*** | -0.014*** | -0.014*** |
|  | (0.001) | (0.001) | (0.001) |
| Constant | 0.752*** | 0.742*** | 0.753*** |
|  | (0.008) | (0.007) | (0.008) |
| Observations | 705,863 | 705,863 | 705,863 |
| $R^2$ | 0.370 | 0.369 | 0.370 |
| Adjusted $R^2$ | 0.369 | 0.369 | 0.370 |
| Residual Std. Error | 2.092 (df = 705839) | 2.092 (df = 705839) | 2.092 (df = 705838) |
| F Statistic | 17,985.710*** (df = 23; 705839) | 17,976.170*** (df = 23; 705839) | 17,239.010*** (df = 24; 705838) |

*Notes:* ***Significant at the 1 percent level.
**Significant at the 5 percent level.
*Significant at the 10 percent level.

*Note:* Some restraint should be taken when interpreting the individual coefficients of model (3), and not as a set. The correlation between BS Errors and BAW Errors is very high at approximately 96%, suggesting an issue of multicollinearity and extreme sensitivity in the estimators, though reasonably low VIFs show that this may not be a particularly big issue.

**Generalized Variance-Inflation Factors**

BS.Errors: 13.9534

BAW.Errors: 13.8788

All Lags: < 2

**Conclusion**

The results of the regression model suggest that the existence of pricing errors tends to have a positive and significant effect on market volume, after controlling for price trends and market momentum. Errors from the Black-Scholes model tend to exhibit a stronger effect on market volume than the Barone-Whaley and Adesi approximation. It is possible then that the Black-Scholes model may be more widely accepted or used in practice, and so any mispricing there may have a greater impact on trading decisions.

As market prices deviate from the models' theoretical prices, a one unit increase in individual pricing errors is expected to increase market volume by 2.7% and 2.4%, respective to BS and BAW. When considering the combined model as a set, an increase in both pricing errors is expected to affect market volume by 5.1% and -2.4%. By constructing a 95% confidence interval of the estimates given mean pricing errors, we can expect that on average approximately 1.7-4.5% of market volume for SPY options traded between 01/02/2015-12/30/2016 could potentially be in the form of systematic noise.

Final Thoughts

As the SPY is one of the deepest options markets in the world, with a highly diversified index and relatively smooth movements, extending this analysis to individual stocks or smaller portfolios may result in much higher rates of systematic noise. The presence of such noise does seem to refute some of the existing literature's views on market efficiency, and shows that there exists at least a small but significant portion of market inefficiency that is driven by fundamental mispricings of theoretical models. Further research can be done to include other pricing models, and their respective estimated effects on trading volume could serve as an indicator to the relative usage of each model in practice.

References

Andersen, Torben, Tim Bollerslev, Francis Diebold, and Heiko Ebens. "The Distribution of Stock Return Volatility." Journal of Financial Economics 61st ser. (2000): 43-76. Web.

Aristotle. Politics. Cambridge, MA: Harvard UP, 2005. Print.

Barone-Adesi, Giovanni, and Robert E. Whaley. "Efficient Analytic Approximation of American Option Values." *The Journal of Finance* 42.2 (1987): 301. Web.

Black, Fischer, and Myron Scholes. "The Pricing of Options and Corporate Liabilities." *Journal of Political Economy* 81.3 (1973): 637-54. Web.

Black, Fischer. "Fact and Fantasy in the Use of Options." *Financial Analysts Journal, 31 (July/August 1975):* 36-41, 61-72. Web.

Fama, Eugene F. "Efficient Capital Markets: A Review of Theory and Empirical Work." *The Journal of Finance* 25.2 (1970): 383. Web.

Rmetrics Core Team, Diethelm Wuertz, Tobias Setz and Yohan Chalabi (2015). fOptions: Rmetrics – Pricing and Evaluating Basic Options. R package version 3022.85.

Samuelson, Paul A. "Rational Theory of Warrant Pricing." *Industrial Management Review* 6.2 (1965): 13. Web.

Tsay, Ruey S. *Introduction to Analysis of Financial Data with R*. Somerset: Wiley, 2013. Print.